\documentstyle[preprint,aps,tighten]{revtex}
\input epsf
\makeatletter
\def\thepage{1-\@arabic\c@page}
\def\@pnumwidth{2em}
\makeatother
\begin{document}
%\twocolumn
\title{\huge \bf  Mechanisms for colour confinement.}
\author{\large Adriano DI GIACOMO}
\address{Dipartimento di Fisica Universit\`a di Pisa and INFN Sezione di Pisa}
\maketitle
\makeatletter
\global\@specialpagefalse
%\def\@oddhead{\REVTeX{} 3.0\hfill Released November 10, 1992}
%\let\@evenhead\@oddhead
%% run page numbers by "chapter", with copyright for first page
\def\@oddfoot{\reset@font\rm\hfill \thepage\hfill
%\ifnum\c@page=1
%  \llap{\protect\copyright{} 1992
%  American Institute of Physics}%
%\fi
} \let\@evenfoot\@oddfoot
\makeatother
%%%%

%%%%
\def\ii{{\rm i}}
\def\QCD{{\it QCD\,}\,}
\def\slashed{{/}\mskip-10.0mu}
\def\Ref#1{(\ref{#1})}% 
\def\diff{{\rm d}}%
\def\ee{{\rm e}}%
\vskip7.0in
Lectures delivered at the CXXX Course of the {\it International School of Physics}
E. Fermi: ``SELECTED TOPICS IN NONPERTURBATIVE QCD''.\\
\hfill Varenna 27 June - 7 July\,\, 1995
\vfill\eject

\section{Introduction}
There exists by now considerable experimental evidence that Quantum
Chromodynamics ({\it QCD}) is the theory of strong interactions.

The {\it QCD} Lagrangean
\begin{equation}
{\cal L} = -\frac{1}{4}{\vec G}_{\mu\nu}{\vec G}^{\mu\nu} +
\sum_f\bar\psi_f\left(\ii \slashed D -
m_f\right)\psi_f\label{eq:1.1}\end{equation}
has the simplicity and the beauty of a fundamental theory.

The notation in \Ref{eq:1.1} is standard. \QCD\  is a gauge theory with gauge
group $SU(3)$, coupled to the quark fields $\psi_f$ which belong to the
fundamental representation $\{3\}$. $f$ indicates flavour species ($u$, $d$,
$c$, $s$, $t$, $b$).

${\vec G}_{\mu\nu}$ is the field strength tensor, belongs to the adjoint
representation $\{8\}$, and has 8 colour components $G^a_{\mu\nu}$.

In terms of the gauge field $A^a_\mu$
\begin{equation}
G^a_{\mu\nu} = \partial_\mu A^a_\nu - \partial_\nu A^a_\mu +
g f^{abc} A_\mu^b A_\nu^c\label{eq:1.2}\end{equation}
$f^{abc}$ are the structure constants of the gauge group, $g$ is the coupling
constant, and $\alpha_s = g^2/4\pi$. ${\vec G}_{\mu\nu}{\vec G}^{\mu\nu}$ in
\Ref{eq:1.1} is a short notation for $\sum_{a=1}^8 G^a_{\mu\nu} G^a_{\mu\nu}$,
\begin{equation}
D_\mu = \partial_\mu - \ii g \vec T \vec A_\mu\label{eq:1.3}\end{equation}
is the covariant derivative. $T^a$ are the generators in the fundamental
representation
\begin{equation}
\left[ T^a,T^b\right] = \ii f^{abc} T^c\label{eq:1.4}\end{equation}
with normalization
\begin{equation}
{\rm Tr}\left( T^a T^b\right) =
\frac{1}{2}\delta^{ab}\label{eq:1.5}\end{equation}
An alternative notation that we will use is
\begin{equation}
A_\mu \equiv \vec A_\mu\cdot\vec T \equiv
\sum_a A^a_\mu T^a\label{eq:1.6}\end{equation}
and
\begin{equation}
G_{\mu\nu} = {\vec G}_{\mu\nu}\cdot \vec T =
\partial_\mu A_\nu - \partial_\nu A_\mu -\ii g\left[ A_\mu,A_\nu\right]
\label{eq:1.7}\end{equation}
In this notation
\begin{equation}
-\frac{1}{4}{\vec G}_{\mu\nu}{\vec G}_{\mu\nu} = -\frac{1}{2}
{\rm Tr}\left(G_{\mu\nu} G^{\mu\nu}\right)\label{eq:1.8}\end{equation}
Tr is the trace on colour indices.

Most of the evidence for the validity of \QCD\  is obtained by probing short
distances. A well known property of the theory is indeed asymptotic
freedom\cite{1,2}: the effective coupling constant vanishes logaritmically at
short distances, or at large momentum transfers. The rate of the decay
$\pi^0\to \gamma\gamma$, which is related to triangle anomaly, tell us that the
number of colours of the quarks is 3. The structure functions of inclusive
lepton - hadron scattering in the deep inelastic region are consistent with the
scale evolution of the coupling constant governed by the perturbative beta
funcion\cite{1,2,3}
\begin{equation}
\frac{\diff \alpha_s(\mu)}{\diff\ln\mu} =
-\frac{\beta_0}{2\pi}\alpha_s^2(\mu) - \frac{\beta_1}{(2\pi)^2}\alpha_s(\mu)^3
+\ldots\label{eq:1.9}\end{equation}
with
\begin{eqnarray}
\beta_0 &=& 11 - \frac{2}{3} N_f\label{eq:1.10}\\
\beta_1 &=& 51 - \frac{19}{3} N_f\label{eq:1.11}\end{eqnarray}
$N_f$ is the number of flavours with $m_f\ll \mu$.
The solution of \Ref{eq:1.9} at one loop is
\begin{equation}
\alpha_s(\mu^2) = \frac{\displaystyle\alpha_s(\Lambda^2)}
{\displaystyle 1 + \beta_0 \alpha_s\ln\frac{\mu^2}{\Lambda^2}}
\label{eq:1.12}\end{equation}
At large distances the coupling constant becomes large (infrared slavery), and
the perturbative expansion sensless.

In addition to that the perturbative series is intrinsecally divergent. One
could think of it as an asymptotic expansion of some funtion of the coupling
constant. Usually the existence of such a function is proved by resummation
techniques, like Borel resummation. For \QCD\  this is problematic\cite{4}.

Alternative methods to perturbation theory must be used to explore large
distances.

A fundamental problem in \QCD\ , which involves large distances is confinement of
colour. At small distances gluons and quarks are visible as elementary
constituents of hadrons. However neither a free quark nor a free gluon has ever
been observed in nature. Upper limits have been established for the production
of quarks in high energy collisions. I will instead quote a limit deduced in
the frame of standard cosmological model\cite{5}. Assuming that the effective
mass of quarks in the primordial quark gluon plasma is of the order of a few
GeV, the ratios $n_q/n_p$ and $n_q/n_\gamma$ of relic quarks and antiquarks to
relic protons and photons can be computed. One finds $n_q/n_p \sim 10^{-12}$ to
be compared to the experimental upper limit $n_2/n_p \leq 10^{-27}$ obtained
from Millikan - like experiments\cite{5'}.

In the language of field theory confinement corresponds to the statement that
asymptotic states are colour singlets. If \QCD\  is the correct theory of hadrons,
then confinement of colour must be built in it, i.e., it must be derivable from
the lagrangean \Ref{eq:1.1}. The problem is then to understand by what
mechanism asymptotic states of \QCD\  are forced to be colour singlets.

\section{Lattice \QCD\ }
The most successful non perturbative approach to \QCD\  is the lattice
formulation\cite{7}. The idea is to regularize the theory by discretizing the
Euclidean space time to a cubic lattice with periodic boundary conditions.

A generic field theory is defined in terms of the Feynman path integral
\begin{equation}
Z[J] = \int{\cal D}\,\varphi\,\exp\left[ - S[\varphi] - \int J\varphi\diff
x\right]\label{eq:2.1}\end{equation}
$Z[J]$ generates all correlation functions.

The integral is performed on the field configurations $\varphi(x)$, which are a
continuous infinity: it is a functional integral. The way to define a functional
integral is to discretize $x$, and then go to the limit in which the density of
discrete values goes to infinity. Lattice corresponds to a step of this
discretizing procedure: on a lattice the number of integration variables is
finite, the integral \Ref{eq:2.1} becomes an ordinary integral and can be
computed numerically, e.g. by Montecarlo methods.

I will not survey in detail the lattice formulation of \QCD\ , but only make a
few general remarks on it\cite{8}.
\begin{itemize}
\item[i)]
On a lattice computations are made from first principles: the theory is
simulated in its full dynamical content.
\item[ii)] To go from lattice to continuum both the ultraviolet cut-off (the
lattice spacing $a$) and the infrared one (the lattice size $L a$) must be
removed. Asymptotic freedom helps to solve the first problem: as the coupling
constant approaches zero the lattice spacing becomes smaller and smaller in
physical units and the coarse structure of the lattice less and less important.
In the usual notation $\beta = 2 N_c/g^2$ ($N_c = 3$ is the number of colours)
\begin{equation}
a(\beta) \mathop\simeq_{\beta\to \infty} \frac{1}{\Lambda_L}
\exp\left(-\frac{\beta}{b_0}\right) \label{eq:2.2}\end{equation}
$\Lambda_L$ is the physical scale of the lattice regularization, the analog of
the usual $\Lambda_{\overline{MS}}$ of the $\overline{MS}$ scheme.

In addition at large $\beta$'s scaling is governed by the perturbative beta
function: this regime is called asymptotic scaling, and helps in extracting
physics from numerical simulations.

To eliminate the influence of the infrared cut-off the lattice size must be
larger than the correlation length.

Notice that the result of numerical simulations are regularized amplitudes,
which have to be properly renormalized to get physics.
\item[iii)] An advantage of the lattice formulation is that the theory is
quantized in a gauge invariant way. The building block is the link $U_\mu(\hat
n)$, which is the parallel transport from the site $\hat n$ to $\hat n + \hat
\mu$
\begin{equation}
U_\mu(\hat n) = \exp\left(\ii g a A_\mu(\hat
n)\right)\label{eq:2.3}\end{equation}
The measure of the Feynman integral is the Haar measure on the gauge group
$\diff U_\mu(n)$: if the group is compact the integration is finite.
\begin{equation}
Z[J] = \int\,\prod_{n,\mu} \diff U_\mu(n) \,\exp\left[-\beta S(U)\right]
\label{eq:2.4}\end{equation}
The action can be written\cite{7}
\begin{equation}
S(U) = -\sum_{n,\mu<\nu} {\rm
Re}\left[\Pi^{\mu\nu}(n)\right]\label{eq:2.5}\end{equation}
As $a\to 0$
\begin{equation}
\beta S(U) \mathop\simeq_{a\to 0} -\frac{1}{4}G_{\mu\nu} G_{\mu\nu} a^4 + {\cal
O}(a^6) \label{eq:2.6}\end{equation}
$\Pi^{\mu\nu}(n)$ is the parallel transport around an elementary square of the
lattice (plaquette) in the plane $\mu,\nu$.

Any other choice of the action which differs from \Ref{eq:2.5} by terms ${\cal
O}(a^6)$ is equally valid, and is expected to produce the same physics in the
limit $a\to 0$. In the language of statistical mechanics these actions belong
to the same class of universality.
\item[iv)] The equilibrium thermodynamics of \QCD\  at temperature $T$ is
obtained by euclidean quantization with periodic boundary conditions in time
(antiperiodic for fermions), in the limit of infinite spacial volume.

On a lattice of size $N_S$ in the three space directions and $N_T$ in the time
direction, the above conditions imply $N_T\ll N_S$ and
\begin{equation}
N_T a(\beta) = \frac{1}{T}\label{eq:2.7}\end{equation}
In \QCD\  a deconfining phase transition takes place at a temperature $T\simeq
150 - 200 {\rm MeV}$: above such temperature hadrons melt into a plasma of
quarks and gluons, and colour is deconfined.

The existence of such a transition has been established by lattice
simulations\cite{9}. By now no clear experimental evidence exists of quark
gluon plasma: however many experiments are on the way\cite{10}.
\end{itemize}
\section{Confinement of colour}
A quantity which characterizes the long range behaviour of the force acting
between a quark-antiquark pair, $Q$ $\bar Q$, is the Wilson loop, $W(C)$.
$W(C)$ is the trace of the parallel transport along a closed path $C$ in space
time
\begin{equation}
W(C) = {\rm Tr}\left\{ {\rm P}\,\exp\ii \int_C A_\mu\diff
x^\mu\right\}\label{eq:3.1}\end{equation}
P indicates ordering along the path $C$.

Any parallel transport from $x_1$ to $x_2$ transforms as a bilocal covariant
under a gauge transformation $\Omega(x)$. If
\begin{equation}
P_{C}(x_1,x_2) = {\rm P}\exp\int_{{}_C x_1}^{x_2}\ii A_\mu\diff x^\mu
\label{eq:3.2}\end{equation}
\begin{equation}
P_{C}(x_1,x_2) \mathop\to_{\Omega(x)} \Omega(x_1) P_{\ell}(x_1,x_2)
\Omega^\dagger(x_2)
\label{eq:3.3}\end{equation}
For a closed loop $x_1 = x_2$, and the trace being cyclic $W(C)$ is gauge 
invariant. Wilson's action \Ref{eq:2.5} is an example of Wilson loop, with
contour $C$ the elementary plaquette.

If as $C$ we take a rectangle of sides $R$ in some space direction and $T$ in
the time direction, let us indicate by $W(R,T)$ the corresponding Wilson loop
(fig. 1).

It can be shown that $W(R,T)$ describes the propagation in time of a $Q$ $\bar
Q$ static pair at distance $R$ and that
\begin{equation}
W(R,T) \mathop\simeq_{T\to \infty} \exp\left[ - T V(R)\right]
\label{eq:3.4}\end{equation}
with $V(R)$ the static potential.

For a confining potential at large distances
\begin{equation}
V(R) = \sigma R \label{eq:3.5}\end{equation}
and
\begin{equation}
W(R,T) \simeq \exp\left[- \sigma T R\right] \label{eq:3.6}\end{equation}
The dependence \Ref{eq:3.6} is known as area-law, and signals confinement.
$\sigma$ is the string tension, whose empirical value is
\begin{equation}
\sigma = \frac{1}{2\pi}\,{\rm GeV^2}\label{eq:3.7}\end{equation}
The area law \Ref{eq:3.6} is observed in numerical simulations at $\beta <
\beta_c$, i.e. below the deconfining transition: above it the string tension
vanishes\cite{11}.

In summary lattice simulations show that colour is confined at low
temperature, and that quarks and gluons can exist as free particles above some
(deconfinement) temperature.

Since the early days of \QCD\  it was suggested\cite{12,13,14} that the mechanism
by which colour is confined could be ``dual type II superconductivity of the
vacuum''. Dual means that the role of electric and magnetic  quantities are
interchanged with respect to ordinary supeconductors. The idea was inspired by
pioneering work on dual models\cite{15}. In ref.\cite{15} it was argued that
the force between a $Q$ $\bar Q$ pair could be produced by field configurations
analogous to Abrikosov flux tubes in superconductivity. This mechanism makes
the energy proportional to the length of the tube as in \Ref{eq:3.5}, and
explains at the same time the string like behaviour of the dual
amplitude\cite{16}. After the advent of \QCD\  the suggestion came naturally that
the flux tubes could be produced by a dual Meissner effect, squeezing the
chromoelectric field acting between $Q \bar Q$ into Abrikosov flux tubes, in
the same way as happens to magnetic field in ordinary superconductors.

Lattice simulations have demonstrated the existence of chromoelectric flux
tubes between $Q \bar Q$ pairs\cite{17}, supporting the mechanism of
confinement by dual superconductivity. A direct test of the mechanism and the
understanding of its origin are of fundamental interest.

\section{Basic superconductivity[18]}
Superconductivity is a Higgs phenomenon. The effective lagrangean\cite{19}
\begin{equation}
{\cal L} = -\frac{1}{4} F_{\mu\nu} F_{\mu\nu} + \left( D_\mu\varphi\right)^\dagger
\left( D_\mu \varphi\right) - V(\varphi) \label{eq:4.1}\end{equation}
with
\begin{equation}
V(\varphi) = -\mu^2 \varphi^\dagger\varphi + \frac{\lambda}{4}\left(\varphi
\varphi\right)^2 \label{eq:4.2}\end{equation}
and
\begin{equation}
D_\mu = \partial_\mu - \ii 2 e A_\mu \label{eq:4.3}\end{equation}
couples the electromagnetic field to a charged scalar field, which describes
Cooper pairs.

${\cal L}$ is invariant under $U(1)$ global rotations
\begin{equation}
\varphi \to \ee^{2\ii e\alpha} \varphi \label{eq:4.4}\end{equation}
If $\mu^2$ in \Ref{eq:4.2} is negative, $V(\varphi)$ has the shape of fig.2.

$\varphi$ acquires a non vanishing {\it vev} $\Phi$ and the $U(1)$ symmetry is
broken down to rotations by angles multiple of $2\pi/q = \pi/e$, $q = 2 e$
being the charge of a pair.

It proves convenient to parametrize the field $\varphi$ by its modulus $\psi$
and its phase $\ee^{\ii q \theta}$
\begin{equation}
\varphi = \psi\,\ee^{\ii q \theta} \label{eq:4.4b}\end{equation}
Under a $U(1)$ transformation by an angle $\alpha$
\begin{equation}
\psi\mathop\to_{U(1)} \psi\qquad  \theta\mathop\to_{U(1)}\theta + \alpha
\qquad A_\mu \to A_\mu + \partial_\mu\alpha\label{eq:4.5}\end{equation}
The covariant derivative of $\varphi$ becomes
\begin{equation}
D_\mu \varphi = \ee^{\ii q \theta}\left[ \partial_\mu \psi + \ii
q\left(\partial_\mu \theta - A_\mu\right)\psi\right]
\label{eq:4.6}\end{equation}
or, putting
\begin{equation}
\tilde A_\mu = A_\mu - \partial_\mu\theta\label{eq:4.7}\end{equation}
\begin{equation}
D_\mu\varphi = \ee^{\ii q \theta}\left[\partial_\mu - \ii q \tilde
A_\mu\right]\psi \label{eq:4.8}\end{equation}
$\tilde A_\mu$ is gauge invariant and
\begin{equation}
\tilde F_{\mu\nu} = \partial_\mu \tilde A_\nu - \partial_\nu\tilde A_\mu =
F_{\mu\nu} \label{eq:4.9}\end{equation}
In terms of the news fields the effective lagrangean becomes
\begin{equation}
{\cal L} = -\frac{1}{4}\tilde F_{\mu\nu}\tilde F_{\mu\nu} + \partial_\mu
\psi\partial_\mu\psi + q^2 \psi^2 \tilde A_\mu\tilde A_\mu - \mu^2\psi^2
-\frac{\lambda}{4}\psi^4\label{eq:4.10}\end{equation}
and the equation of motion for the electromagnetic field
\begin{equation}
\partial_\mu \tilde F_{\mu\nu} + q^2 \psi^2 \tilde A_\nu
\label{eq:4.11}\end{equation}
Neglecting vacuum fluctuations, $\psi^2 = \bar\psi^2$ (its {\it vev}) and
\Ref{eq:4.11} reads
\begin{equation}
\partial_\mu \tilde F_{\mu\nu} + m^2 \tilde A_\nu = 0
\label{eq:4.12}\end{equation}
with
\begin{equation}
m = \sqrt{2}\,q \bar\psi\label{eq:4.12a}\end{equation}
The parametrization \Ref{eq:4.7} for the field corresponds to the well known
fact that, in the Higgs phenomenon, the phase of the Higgs field supplies the
longitudinal component of the photon, when it becomes massive.

In the gauge $A_0 = 0$ a static configuration, $\partial_0\vec A = 0$,
$\partial_0\varphi = 0$ implies $E_i = F_{0i} = 0$ and \Ref{eq:4.12} for the
space components reads
\[ \partial_i F_{ij} + m^2 \tilde A_j = 0\]
or
\begin{equation}
\vec\nabla\wedge\vec H + m^2\vec{\tilde A} = 0 \label{eq:4.14}\end{equation}
$\vec j = m^2 \vec{\tilde A}$ is known as London current: physically it implies
the existence of a steady current at zero $\vec E$, or, since $\rho \vec j =
\vec E$, zero resistivity.

Taking the curl of both sides of \Ref{eq:4.14} we obtain
\begin{equation}
\nabla^2\vec H - m^2 \vec H = 0\label{eq:4.15}\end{equation}
Eq.\Ref{eq:4.15} means that $\vec H$ penetrates inside the superconductor by a
length $\lambda_1 = 1/m$ and is otherwise expelled from the bulk of it: this
fact is known as Meissner effect.

The key parameter, both for producing zero resistivity and Meissner effect is
$\Phi = \langle |\varphi|\rangle\neq 0$ being $m^2 = 2 q^2|\Phi|^2$
\Ref{eq:4.12a}.

$\Phi\neq 0$ indicates that $U(1)$ symmetry is spontaneously broken. Indeed if
the vacuum $|0\rangle$ were $U(1)$ invariant, it would be
\begin{equation}
U(1) |0\rangle = \ee^{\ii q_0 \alpha}|0\rangle \label{eq:4.16}\end{equation}
and, for any operator ${\cal O}$
\begin{equation}
\langle 0| {\cal O} |0\rangle =
\langle 0| U(1)^\dagger{\cal O} U(1) |0\rangle \label{eq:4.17}\end{equation}
If ${\cal O}$ carries charge $q$, $U^\dagger {\cal O} U = \ee^{\ii q\alpha} {\cal
O}$ and eq.\Ref{eq:4.17} gives
\begin{equation}
\langle 0| {\cal O} |0\rangle = \ee^{\ii q\alpha} 
\langle 0| {\cal O} |0\rangle \label{eq:4.18}\end{equation}
which implies either $q=0$ or $\langle 0| {\cal O} |0\rangle = 0$. A non zero
{\it vev} of any charged operator, like $\varphi$, implies that vacuum is not
invariant under
$U(1)$, which is a symmetry of ${\cal L}$, i.e. that $U(1)$ symmetry is
spontaneously broken.

A non zero {\it vev} of any charged operator signals superconductivity: the
effective lagrangean for that operator, due to the universal coupling to
$A_\mu$, does indeed generate a mass for the photon. The ground state of a
superconductor is a superposition of states with different charges (Cooper pair
condensation).

In the above equations two scales of length appear: $\lambda_2 = 1/\mu$, the
inverse mass or correlation length of the Higgs field, and $\lambda_1 = 1/m$,
the inverse mass of the photon or penetration depth of the magnetic field. A
superconductor is called type II if $\lambda_1 \gg \lambda_2$, otherwise it is
named type I. The formation of Abrikosov flux tubes at intermediate values of
the external field $\vec H$, is energetically favorable in type II
superconductors. In type I there is complete Meissner effect below some value
of $H_c$ of $H$, and complete penetration of it above $H_c$, with destruction
of superconductivity.
\section{Monopoles}
\subsection{Generalities on monopoles}
The difficulty in the construction of operators with non zero magnetic charge
in terms of the gauge fields $A_\mu$, stems from the fact that monopole
configurations have non trivial topology. Here we shall briefly review the
definition and the classification of monopoles. For a more detailed treatment
we refer to re.\cite{20}, from which most of what we say here is extracted.

The most general form of Maxwell's equation in the presence of both electric
($j_\mu$) and magnetic ($j^M_\mu$) current is
\begin{eqnarray}
\partial_\mu F_{\mu\nu} &=& j_\nu \label{eq:5.1}\\
\partial_\mu F^*_{\mu\nu} &=& j^M_\nu \nonumber
\end{eqnarray}
where $F^*_{\mu\nu} = \frac{1}{2}\varepsilon_{\mu\nu\rho\sigma} F^{\rho\sigma}$
is the dual of $F_{\mu\nu}$. Free field equations ($j_\nu = j^M_\nu = 0$) are
trivially invariant under the group of transformations
\begin{eqnarray}
F_{\mu\nu} &\to& \cos\theta F_{\mu\nu} + \sin\theta F^*_{\mu\nu} 
\label{eq:5.2}\\
F^*_{\mu\nu} &\to& \cos\theta F^*_{\mu\nu} - \sin\theta F_{\mu\nu}
\nonumber
\end{eqnarray}
or
\begin{eqnarray}
\vec E &\to& \vec E \cos\theta + \vec H \sin\theta \label{eq:5.3}\\
\vec H &\to& \vec H \cos\theta - \vec E\sin\theta\nonumber\end{eqnarray}
In particular this holds for $\theta = \pi/2$ or
\begin{equation}
\vec E\to \vec H\qquad \vec H \to -\vec E\label{eq:5.4}\end{equation}
\Ref{eq:5.4} is known as duality transformation.

No particles with magnetic charges have ever been observed in nature, in spite
of the many attempts to detect them. As a consequence Maxwell's equations are
usually written
\begin{mathletters}
\begin{eqnarray}
\partial_\mu F^{\mu\nu} &=& j^\nu\label{eq:5.5a}\\
\partial_\mu F^{*\mu\nu} &=& 0\label{eq:5.5b}\end{eqnarray}
\end{mathletters}
The general solution of eq.\Ref{eq:5.5b} is
\begin{equation}
F_{\mu\nu} = \partial_\mu A_\nu - \partial_\nu A_\mu\label{eq:5.6}\end{equation}
with $A_\mu$ an arbitrary vector field. For any $A_\mu$ eq.'s \Ref{eq:5.5b}are
identically satisfied. They are indeed known as Bianchi identities. The very
possibility of introducing $A_\mu$ relies on the absence of magnetic charges,
eq.'s \Ref{eq:5.5b}. Explicitely they read
\begin{equation}
{\rm div}\vec H = 0\qquad \frac{\partial \vec H}{\partial t} -
\vec\nabla\wedge\vec E = 0\label{eq:5.7}\end{equation}
If a monopole exists, and we insist in describing the system in terms of
$A_\mu$, the monopole must be viewed as one end of a long, infinitely thin
solenoid, bringing the magnetic flux to infinity, in order to satisfy the first
of eq.\Ref{eq:5.7}\cite{21}(fig. 3).

The solenoid is known as Dirac string. In order to make this string physically
invisible, the parallel transport of any charged particle on any closed path
around it must be equal to 1, or
\[ 2\pi n = e\int_C \vec A\diff\vec x = e \Phi(H)\]
where $\Phi$ is the flux of the magnetic field across the string. On the other
hand by construction $4\pi Q_M = \Phi$ hence
\begin{equation}
Q_M = \frac{n}{2 e}\label{eq:5.8}\end{equation}
This is the celebrated Dirac quantization for the magnetic charge.

In conclusion a monopole field can still be described in terms of $A_\mu$,
provided eq.\Ref{eq:5.8} is satisfied, at the price of introducing a nontrivial
topology. The argument can be generalized to non abelian gauge theories and to
a generic distribution of charges contained in a finite region of space. The
idea is to look at the multipole expansion of the field at large distances, and
to allow for a nonzero magnetic monopole component. We shall restrict for
simplicity to configurations with zero electric field at some time, say $t=0$:
a sufficient condition for that is that in some gauge
\[ A_0(t) = - A_0(-t),\quad \vec A(t)=\vec A(-t)\]
or
\[ A_0 = 0\qquad \frac{\partial\vec A}{\partial t} = 0\]
at $t=0$. Then
\[ F_{0i} = \partial_0 A_i - \partial_i A_0 - \ii g\left[ A_o,A_i\right] = 0\]
Let us denote by $\vec A \equiv (A_r,A_\theta,A_\varphi)$ the components of
$\vec A$ in polar coordinates. $A_r$ can be made zero by a time independent
gauge transformation, which does not affect $A_0$. Let
$\Lambda(r,\theta,\varphi)$ be the parallel transport to infinity along the
radius
\begin{equation}
\Lambda = {\rm P}\exp\left(\ii\int_r^\infty
A_r(r,\theta,\varphi)\,\diff\varphi\right)\label{eq:5.9}\end{equation}
$\Lambda$ is unitary, $\Lambda^\dagger = \Lambda^{-1}$. Moreover
\begin{equation}
\frac{\partial \Lambda}{\partial r} + \ii A_r\Lambda =
0\label{eq:5.10}\end{equation}
Multiplying by $\Lambda^\dagger$ to the left \Ref{eq:5.10} gives
\begin{equation}
A'_r = -\ii \Lambda^\dagger\frac{\partial\Lambda}{\partial r} +
\Lambda^\dagger A_r \Lambda = 0\label{eq:5.11}\end{equation}
$A'_r$ is in fact the gauge transformed of $A_r$ under $\Lambda$. We will be
interested at the behaviour of the field at distances $r > R$ and therefore we
do not worry about singularities at $r=0$.
We are thus left with $\vec A = (0,A_\theta,A_\varphi)$, and we look for a
configuration behaving as $1/r$ as $r\to\infty$ or
\[ 
\vec A =
(0,\frac{a_\theta(\theta,\varphi)}{r},
\frac{a_\varphi(\theta,\varphi)}{r})\]
Again we can make $a_\theta = 0$ by a procedure similar to the one used to make
$A_r=0$. We operate a gauge transformation, independent of $t$ and $r$
\begin{equation}
\Lambda' = {\rm P}\exp\left(\ii\int_0^\theta a_\theta(\theta,\varphi) \diff\theta\right)
\label{eq:5.12}\end{equation}
We are then left with $a_\varphi$ alone and the only non vanishing component of
$F^{\mu\nu}$ is $F^{\theta\varphi} = \partial_\theta a_\varphi$.
The field equations, outside the space occupied by matter ($r>R$) read
\begin{equation}
\partial_\mu\sqrt{g} F^{\mu\nu} + \left[A^\mu,\sqrt{g} F^{\mu\nu}\right] = 0
\label{eq:5.13}\end{equation}
$g$ is the determinant of the metric tensor and $\sqrt{g} = 1/r^2\sin^2\theta$.
For $\nu=\varphi$ eq.\Ref{eq:5.13} gives
\begin{equation}
\partial_\theta\frac{1}{\sin\theta}\partial_\theta a_\varphi = 0
\label{eq:5.14}\end{equation}
which has the general solution
\[
a_\varphi = Q(\varphi)(a+b\cos\theta) \]
If we want no singularity at the north pole $a_\varphi(0) = 0$ and
\begin{equation}
a_\varphi = Q(\varphi)(1-\cos\theta) \label{eq:5.15}\end{equation}
The equation with $\nu=\theta$ reads
\begin{equation}
\partial_\varphi \sqrt{g} F^{\varphi\theta} + \left[a^\varphi,\sqrt{g}
F^{\varphi\theta}\right] = 0 \label{eq:5.16}\end{equation}
The term with the commutator cancels and the net result is
\begin{equation}
\partial_\varphi Q(\varphi) = 0 \label{eq:5.16b}\end{equation}
or $Q={\rm cost}$. The non abelian monopole field is the abelian field times a
constant matrix $Q$: in eq. \Ref{eq:5.14}, \Ref{eq:5.16} the term
with the commutator which signals the non abelian nature of the gauge group has
disappeared.

By our choice the Dirac string lies along the axis $\theta=\pi$.

If we had chosen the string along the axis $\theta=0$ we would obtain
\[ a_\varphi = - Q(1+\cos\theta)\]
The two configuration differ by a gauge transformation
\[ U = \ee^{\ii 2 Q\varphi}\]
If we demand $U$ to be single valued
\begin{equation}
\exp(\ii 4\pi Q) = 1\label{eq:5.17}\end{equation}
which is the Dirac quantization condition if we keep in mind that in our
notation $A_\mu$ incorporates the coupling constant $g$. Eq.\Ref{eq:5.17} gives
\begin{equation}
g Q_{ii} = \frac{m_i}{2} \label{eq:5.18}\end{equation}
with $m_i$ an integer.

To summarize a monopole configuration is identified by a constant diagonal
matrix $Q$ of the algebra, with integer or half integer eigenvalues, up to a
a gauge transformation. This identification is known as GNO (Goddard, Nuyts,
Olive) classification\cite{22}. 

In $SU(N)$ $Q$ is traceless, and is therefore identified by $N-1$ half integer
eigenvalues: they are $N-1$ charges corresponding to the residual $U(1)^{N-1}$
gauge invariance, under the transformations which leave a matrix $Q$ diagonal.

For configurations containing many monopoles it can be shown that by use of
gauge transformations the matrix $Q$ can be constructed, which is the sum of
the $Q_i$'s describing the single monopoles.

What matters is the topology, specifically the homotopy structure of the
$(N-1)$ $U(1)$'s obtained by GNO construction\cite{22}. It can also be shown
that the stable monopole configurations are possible only if the first homotopy
group of the gauge group, $\Pi_1(G)$, is non trivial\cite{23}, i.e. if the group $G$
is non simply connected.

$\Pi_1(SU(N))$ is always trivial: there exist no stable classical monopole
configurations unless the symmetry is broken down to a subgroup $H$ with non
trivial homotopy, like $U(1)$.
\subsection{'t Hooft - Polyakov monopoles}
A well known example of classical monopole configuration is the 't Hooft -
Polyakov\cite{24,25} monopole. The theory\cite{26} is an $SU(2)$ gauge theory
coupled to a scalar field in the adjoint representation $\vec \varphi$: since
all fields have integer isospin, the gauge group is $SO(3)$ or $SU(2)/Z_2$. The
lagrangean is
\begin{equation}
{\cal L} = \frac{1}{2}( D_\mu\vec\varphi)(D_\mu\varphi) - \frac{1}{4}
\vec G_{\mu\nu}\vec G_{\mu\nu} - V(\vec\varphi^2)\label{eq:v2.1} \end{equation}
where $D_\mu\vec \varphi = \partial_\mu\vec \varphi - g\vec
A_\mu\wedge\vec\varphi\quad$ is the covariant derivative and
\begin{equation}
V(\vec\varphi^2) = \frac{\mu^2}{2}\vec\varphi^2 +
\frac{\lambda}{4}(\vec\varphi^2)^2 \label{eq:v2.2}\end{equation}
is the most general potential invariant under the gauge group and compatible
with renormalizability.

If $\mu^2 < 0$, $\vec \varphi$ acquires a nonvanishing {\it vev\,}
$\langle\vec\varphi\rangle$, and the symmetry group breaks down to $U(1)$,
which is the group of isospin rotations around the direction of
$\langle\vec\varphi\rangle$. An explicit static solution of the equations of
motion with finite energy can then be constructed.

With a special choice of the gauge the solution is a ``hedgehog'':
\begin{equation}
A^j_\mu = -\varepsilon_{\mu jk}\frac{r_k}{r^2} A(r)\qquad
\frac{\varphi^i}{\langle\vec\varphi\rangle} =
\frac{r^i}{r} F(r) \label{eq:v2.3}\end{equation}
(the upper index is an isospin index).

At large distances $A(r)\to 1$, $F(r)\to 1$. By continuity $\vec\varphi$ must
vanish at some point, which is identified with the location of the monopole.

The solution can be gauge rotated to a gauge (unitary gauge) in which
$\vec\varphi$ is oriented along the 3-axis $\vec\varphi = (0,0,\Phi)$: this
gauge transformation is regular everywhere except in the points where
$\vec\varphi = 0$, and coincides with the construction of sect.~1. In fact the
configuration \Ref{eq:v2.3} $A_0=0$, $A_r=0$
\begin{eqnarray}
A_\theta &=& \frac{1}{g}\vec\sigma(\vec n_\perp\wedge\vec \nu)\label{eq:v2.4}\\
A_\varphi &=& \frac{1}{g}\left[(\vec\nu\cdot\vec n)\vec n -
\vec\nu\right]\vec\sigma\nonumber
\end{eqnarray}
with $\vec n_\perp = (\cos\varphi,\sin\varphi,0)$; $\vec\nu = (0,0,1)$; $\vec n
= \vec r/r$. $\vec n_\perp\wedge\vec \nu$ is $\theta$ independent. Therefore the
gauge transformation $\Lambda_\theta$ which makes $A_\theta=0$ can be written
\[ \Lambda_\theta \equiv {\rm P}\exp\int_0^\theta g A_\theta\diff\theta=
\exp \ii\theta\vec\sigma(\vec n_\perp\wedge\vec \nu)\]
It is easy to check that
\begin{mathletters}
\begin{eqnarray}
\Lambda_\theta^\dagger\vec n\cdot\vec\sigma\Lambda_\theta &=& \sigma_3 
\label{eq:v2.5a}\\
\Lambda^\dagger_\theta A_\theta \Lambda_\theta + \ii
\Lambda_\theta^\dagger\partial_\theta \Lambda_\theta &=& 0
\label{eq:v2.5b}\\
\Lambda_\theta^\dagger A_\varphi \Lambda\theta + \ii \Lambda_\theta^\dagger
\partial_\varphi \Lambda_\theta &=& \sigma_3\frac{(1-\cos\theta)}{2}
\label{eq:v2.5c}\end{eqnarray}
\end{mathletters}
The gauge transformation which makes $\varphi$ diagonal is called an abelian
projection. The abelian projection coincides with the GNO construction. The
matrix $Q$ of eq.\Ref{eq:5.18} is in this model $Q = 2\sigma_3$, corresponding
to a monopole of charge 2 Dirac units.

A gauge invariant field strength $F_{\mu\nu}$ can be defined\cite{24}
\begin{equation}
F_{\mu\nu} = \hat\varphi\cdot\vec G_{\mu\nu} - \frac{1}{g}\hat\varphi\cdot
\left( D_\mu\hat\varphi \wedge D_\nu \hat\varphi\right)
\label{eq:v2.6}\end{equation}
whit $\hat\varphi=\vec\varphi/|\vec\varphi|$.

Similarly we can define
\begin{equation}
B_\mu = \hat\varphi\cdot \vec A_\mu \label{eq:v2.7}\end{equation}
$B_\mu$ is not gauge invariant, since $\vec A_\mu$ is not covariant under gauge
transfromations. In fact $A_\mu\to U^\dagger A_\mu U + \ii
U^\dagger\partial_\mu U$. The identity holds
\begin{equation}
F_{\mu\nu} = (\partial_\mu B_\nu - \partial_\nu B_\mu) - \frac{1}{g}
\hat\varphi\left( \partial_\mu\hat\varphi \wedge \partial_\nu \hat\varphi\right)
\label{eq:v2.8}\end{equation}
The two terms in eq.\Ref{eq:v2.8} are not separately gauge invariant; only their
sum is. After abelian projection the second term drops and $F_{\mu\nu}$ is an
abelian gauge field, with vector potential $B_\mu$
\begin{equation}
F_{\mu\nu} = \partial_\mu B_\nu - \partial_\nu B_\mu\qquad
\label{eq:v2.9}\end{equation}
For the solution of the form \Ref{eq:v2.3} $F_{\mu\nu}$ as defined by
eq.\Ref{eq:v2.8} obeys free Maxwell equations, except in the point where $\vec
\varphi = 0$
\[ \partial_\mu F^{\mu\nu} = 0\]
At large distances (eq.\Ref{eq:v2.3})
\[ \vec E = 0\qquad \vec H = \frac{1}{g}\frac{\vec r}{r^3}\]
$F_{\mu\nu}$ for this solution is the field of a pointlike Dirac monopole. Note
that in the hedgehog gauge of \Ref{eq:v2.3} only the second term of eq.\Ref{eq:v2.8}
contributes to the magnetic field.
\subsection{The abelian projection in QCD}
In \QCD\  there is no Higgs field. However any operator $\Phi(x)$ in the adjoint
representation can act as an effective Higgs field\cite{28}, and can be used to
define monopoles. In what follows we shall refer to $SU(2)$ gauge group for the
sake of simplicity: the extension to $SU(3)$ is trivial.

In the notation of previous section 
\begin{equation}
\Phi(x) = \vec\Phi(x)\vec\sigma  \label{eq:v3.1}\end{equation}
The abelian projection is by definition a gauge transformation which
diagonalizes $\Phi(x)$: like in the 't Hooft - Polyakov's monopole
configuration analized in sect. 3 ,it is singular when $\Phi(x) = 0$. Those
zeros will be then world lines of monopoles.

A gauge invariant field strength
\begin{equation}
F_{\mu\nu} = \hat\varphi\cdot\vec G_{\mu\nu} - \frac{1}{g}\hat\varphi\cdot
\left( D_\mu\hat\varphi \wedge D_\nu \hat\varphi\right)
\label{eq:v3.2}\end{equation}
can be defined, which in fact is the field of a Dirac monopole in the
neighbouring of a zero. Again, putting $B_\mu = \hat\Phi\cdot\vec A_\mu$
\begin{equation}
F_{\mu\nu} = (\partial_\mu B_\nu - \partial_\nu B_\mu) - \frac{1}{g}
\hat\varphi\left( \partial_\mu\hat\varphi \wedge \partial_\nu \hat\varphi\right)
\label{eq:v3.3}\end{equation}
In a hedgehog gauge ($\hat\Phi = \hat r$) the first term in
\Ref{eq:v3.3} does not contribute to monopole charge: the second term carries a
net flux of magnetic field $4\pi/g$.
After abelian projection instead only the first term is different from zero.

The classical construction of GNO can now be given a quantum mechanical
extension.

Indeed, in each configuration appearing in the Feynman integral, monopoles will
be located in the zeros of the classical version of $\Phi(x)$. The abelian
projection will identify the GNO matrix defining the monopole charges: monopole
charges will be additive, the field strength beeing asymptotically abelian, and
the field equations linear.

Viceversa given a classical configuration with pointlike monopoles a patching
of the single monopole configurations can be performed by gauge transformations,
in such a way that the monopole charges are additive: the basic fact is that
monopole charge is a topological property, which can be described by homotopy,
and a group product in the set of paths around the Dirac strings of the
different monopoles can be defined, such that the winding numbers are additive.
As a consequence the GNO matrix for the configuration will be the sum of
the matrices for the single monopoles. What really matters is the charge and
the location of the monopoles. Any operator in the adjoint representation which
is zero in the location of the monopoles, and is diagonal with $Q$ will
identify an abelian projection which makes $Q$ diagonal.

Defining such operator will allow to label monopoles in different classical
configurations which enter the Feynman integral.

In conclusion all monopoles are $U(1)$  monopoles. A monopole species is
identified by an operator in the adjoint representation. Notice that the monopole
charge coupled to the magnetic field of eq.\Ref{eq:v3.2} is invariant under the
gauge group: such charges can condense in the vacuum without breaking the gauge
symmetry

Notice also that if $SU(N)$ gauge symmetry is not broken, monopoles are unstable.
\subsection{Monopoles and confinement in \QCD\ }
Any operator ${\cal O}(x)$ in the adjoint representation defines a monopole
species by abelian projection: of course different choices for ${\cal O}(x)$
will define different monopole species: the number and the location of
monopoles will indeed depend on the zeros of ${\cal O}(x)$. Dual
superconductivity means condensation of monopoles in the vacuum.

Popular choices for the abelian projection correspond to the following choices
for ${\cal O}(x)$
\begin{itemize}
\item[1)]
\phantom{a}\par\noindent
\begin{equation}
{\cal O}(x) = P(x) \label{eq:v4.1}\end{equation}
$P(x)$ beeing the Polyakov line
\begin{equation}
P(\vec x,x_0) = {\rm P}\exp\left[ \ii\oint A_0(\vec x,t)\diff t\right]
\label{eq:v4.2}\end{equation}
$P$ is the parallel transport in time on the closed path from $x_0$ to $+\infty$
and back from $-\infty$ to $x_0$. $P(x)$ transforms covariantly under the
adjoint representation. On a lattice, due to periodic b.c., $P(x)$ is defined as
\[ P(\vec n,n_0) = \prod_{i=1}^{N_T} U_0(\vec n,n_0 + i\nu_0)\]
$N_T$ beeing the number of links in the time direction.
\item[2)] 
\phantom{a}\par\noindent
\[ {\cal O}(x) = F_{ij}\]
$F_{ij}$ any component of the field strength tensor.
\item[3)] Maximal abelian\cite{29}. This projection is defined on the lattice by the
condition that the quantity
\[ M = \sum_{n,\mu}{\rm Tr}\,\left\{ U_\mu(n)\sigma_3 U^\dagger_\mu(n)
\sigma_3\right\}\]
be maximum with respect to gauge transformation $\Omega(n)$. In formulae:
\[ 0 = \frac{\delta M}{\delta \Omega(m)} =
\frac {\delta }{\delta \Omega(m)} \sum_{n,\mu}
\left\{ \Omega(n) U_\mu \Omega^\dagger(n+1) \sigma_3 \Omega(n+1) U^\dagger_\mu
\Omega^\dagger(n)\sigma_3\right\}\]
The operator ${\cal O}(x)$ in this case is known only in the gauge where $M$ is
maximum, but non in its covarian form. In that gauge
\[ {\cal O}(n) =
\sum_\mu U^\dagger_\mu(n) \sigma_3 U_\mu(n) +
U_\mu(n-\mu) \sigma_3 U^\dagger_\mu(n-\mu) \]
\end{itemize}
The question relevant to physics is: what monopoles do condense in the \QCD\ 
vacuum to produce dual superconductivity, if any.

As discussed in the introduction this question is better adressed on the
lattice. From the point of view of physics a possibility is that all monopoles,
defined by different abelian projections, do condense in the vacuum\cite{28}. Is this
the case?

An answer to the above question is possible if we develop a tool to 
directly detect dual supeconductivity. Understanding the problem in $U(1)$ gauge
theory will be sufficient, since the problem reduces in any case to $U(1)$
after abelian projection.
\section{Detecting dual superconductivity}
To detect dual superconductivity one can  use a phenomenological
approach, which consists in the detection of a London current: this approach is
discussed in the lectures of D. Haymaker, in this course\cite{30}.

An alternative method, which will be discussed here, is to detect a
nonvanishing {\it vev} of a quantity with nonzero magnetic charge. In the next
section we shall construct an operator with non zero magnetic charge, which
will be used as a probe of dual superconductivity.

\subsection{$U(1)$ gauge group.\label{secu1}}
Pure $U(1)$ gauge theory in the continuum formulation is a theory of free
photons. On the lattice, the building block of the theory beeing the link
$U_\mu(n) = \exp(\ii e A_\mu(n) a)$, and the action beeing the plaquette
(Wilson's action) interactions exist to all orders in $e$. Putting
$\beta=1/e^2$ a value $\beta_c$ exists, $\beta_c \simeq 1.0114(2)$ such that
for $\beta > \beta_c$ the system is made of free photons. For $\beta < \beta_c$
instead the interaction is strong and Wilson loops obey the area law, which
implies confinement of electric charge.

In the confined phase monopoles condense in the vacuum, which behaves as a
superconductor. This has been rigorously proven in ref.\cite{31}, but only for a
special form of the action (the Villain action), by showing that a magnetically
charged operator exists, whose {\it vev} is different from zero. In
ref.\cite{32} the construction has been extended to a generic form of the
action.

The basic idea is the well known formula for translation
\begin{equation}
\ee^{\ii p a} | x\rangle = |x+a\rangle \label{eq:6.1}\end{equation}
The analog of \Ref{eq:6.1} for a gauge field is
\begin{equation}
\exp\left[\ii\int\vec\Pi(\vec x)\cdot\vec b(\vec x)\diff^3 x\right]
|\vec A(\vec x)\rangle = |\vec A(\vec x) + \vec b(\vec x)\rangle
\label{eq:6.2}\end{equation}
where $\vec A(\vec x)$ play the role of $q$ coordinates, $\vec\Pi(\vec x) =
\partial{\cal L}/\partial \dot{\vec A(\vec x)}$ play the role of conjugate
momenta, and $|\vec A(\vec x)\rangle$ is a state in the Schr\"odinger
representation.
In the continuum
\begin{equation}
\Pi^i(\vec x,t) = F^{0i}(\vec x,t)\label{eq:6.3}\end{equation}
If we choose $\vec b$ as the vector potential describing the field of a
monopole of charge $m/2g$, with the Dirac string in the direction $\vec n$
\begin{equation}
\vec b(\vec x,\vec y) = - \frac{m}{2 g}\frac{\vec n\wedge\vec r}{r(r - \vec
r\cdot\vec n)}\qquad \vec r = \vec x - \vec y \label{eq:6.4}\end{equation}
then
\begin{equation}
\mu(\vec y,t) = \exp\left\{\ii\int\diff^3 x\,\vec \Pi(\vec x,t)\vec b(\vec
x,\vec y)\right\} \label{eq:6.5}\end{equation}
creates a monopole of charge $m$ Dirac units at the site $\vec y$ and time $t$.

The magnetic charge operator is
\[ Q = \int\diff^3x\,\vec \nabla\vec H = \int\diff^3x
\vec\nabla\cdot(\vec\nabla\wedge\vec A)\]
The commutator $[Q(t),\mu(\vec y,t)]$ can be easily evaluated, using the basic
formula
\[ \left[ \ee^{\ii p a}, q\right] = q + a \]
giving
\[[Q(t),\mu(\vec y,t)] = \int{\diff^3}x \vec\nabla\cdot\left(\vec\nabla\wedge\vec
b(\vec x,\vec y)\right)\mu(\vec y,t) = \frac{m}{2 g}\mu(\vec y,t)\]
In eq.\Ref{eq:6.5} the Dirac string potential has been subtracted from $\vec b(\vec
x,\vec y)$.

We will compute the {\it vev}
\[\tilde \mu = \langle 0| \mu(\vec y,t)|0\rangle \]
as a possible probe of spontaneous symmetry breaking of dual $U(1)$, i.e. as a
probe of dual superconductivity.

On the lattice  $\Pi^i = \frac{1}{e}{\rm Im} \Pi^{0i}$ and the
obvious transcription for $\mu$ is, after Wick rotation to Euclidean space,
\begin{equation}
\mu(\vec n,n_0) = \exp\left[ - \beta\sum_{\vec n'}{\rm Im} \Pi^{0i}(\vec n',n_0)
b^i(\vec n' -\vec n)\right] \end{equation}
$ b^i$ is the discretized version of $\vec b$, and the factor $\beta$ comes
from the $1/g$ of monopole charge and $1/g$ of the field. Of course $\mu$
depends on the gauge choice for $\vec b$. A better definition, which makes
$\mu$ independent of the choice of the gauge for $\vec b$ is
\begin{equation}
\mu(\vec n,n_0) = \exp\left\{ - \beta\sum_{\vec n'}{\rm Re} 
\left[\Pi^{0i}(\vec n',n_0)
(\ee^{b^i}-1)\right]\right\}\label{eq:mu1} \end{equation}
which coincides with the previous equation for small values of $b_i$. The line
of sites corresponding to the location of the string must be subtracted: $\vec n'$
in eq.\Ref{eq:mu1} runs on all sites except the string.

When computing $\langle \mu\rangle$ whith the Feynman integral
\begin{equation}
\langle \mu \rangle = \frac{\displaystyle\int \prod [\diff
U_\mu]\ee^{-S} \mu(\vec n,n_0)} {\displaystyle\int \prod [\diff
U_\mu]\ee^{-S}}\label{eq:7.b1}\end{equation} it can be easily shown that a gauge
transfromation on $\vec b$, is reabsorbed by the invariance of the Haar measure of
integration, if the compactified form \Ref{eq:mu1} is used.

Eq.\Ref{eq:7.b1} coincides with the construction of ref.\cite{31} when the action
has the Villain form.

A direct measurement of $\langle\mu\rangle$ gives problems with the
fluctuations: indeed $\mu$ is the exponent of a quantity which is roughly
proportional to the volume, and therefore has fluctuations of the order
$V^{1/2}$. It is a well known fact in statistical mechanics that such
quantities are not gaussian-distributed, and that the width of their
fluctuations does not decrease with increasing statistics: the same problem
occurs with the numerical determination of the partition function. Therefore,
instead of $\langle\mu\rangle$ we will measure the quantity
\begin{equation}
\rho(\beta) = \frac{\diff}{\diff \beta}\ln\langle\mu\rangle
\label{eq:7.b2}\end{equation} which is the analog of the internal energy in the
case of partition function.
$\langle \mu\rangle$ can then be reconstructed from eq.\Ref{eq:7.b2}, by use of
the boundary condition $\langle \mu\rangle = 1$ ($\beta=0$), obtaining
\begin{equation}
\langle\mu\rangle = \exp\int_0^\beta \rho(\beta')\diff \beta' \end{equation}
If $\langle\mu\rangle$ has a shape like in fig.4.

The expected behaviour of $\rho$ is depicted in fig.5.

we expect a sharp negative peak around $\beta_c$. Notice that $\langle\mu\rangle$ is
an analitic function of $\beta$ for a finite lattice: so it cannot be exactly
zero above $\beta_c$. Only in the infinite volume limit singularities can
develop in the partition function and $\langle\mu\rangle$ can be zero in the
ordered phase without being identically zero.

The result of numerical simulations with Wilson action are shown in fig.6.

It is known that the transition at $\beta_c$ is weak first order\cite{33}: this
means that, approaching $\beta_c$ from below the correlation length increases
as in a second order transition, up to a certain value $\tilde \beta ,
\beta_c$, where this increase stops. For values of $\beta < \tilde \beta$ a
finite size analysis can be performed. $\langle\mu\rangle$ in principle depends
on $\beta$, on the lattice spacing and on the lattice size $L$, and for a
finite lattice is an analytic function of $\beta$. Since the correlation length
$\xi$ is proportional to $(\beta_c-\beta)^{-\nu}$ for $\beta \leq \tilde \beta$
the $\beta$ can be traded with $\xi$ and $\nu$.

Finite size scaling occours when $a/\xi \ll 1$, $a/L\ll 1$ are not relevant, and
\begin{equation}
\langle\mu\rangle=\Phi(L^{1/\nu}(\beta-\beta_c),L)\end{equation}
since $\xi^{-1/\nu}\sim (\beta_c - \beta)$.

As $\beta$ approaches $\tilde\beta_c$ $\langle\mu\rangle$ must vanish in the limit
$L\to \infty$ and
\begin{equation}
\rho = \frac{\diff}{\diff\beta}\ln\langle\mu\rangle \simeq
L^{-1/\nu}\frac{\displaystyle
\Phi'(L^{1/\nu}(\beta_c-\beta))}{\displaystyle\Phi}\end{equation}
$L^{-1/\nu}\rho$ is a universal function of $L^{1/\nu}(\beta_c-\beta)$. 
The quality of this scaling is shown in fig.7.

If at infinite volume $\langle\mu\rangle\simeq (\beta_c-\beta)^{\delta}$ the
index $\delta$ and $\beta_c$ can be extracted from that universal function.
By a best fit we obtain
\begin{mathletters}
\begin{eqnarray}
\nu &=& 0.29 \pm 0.1\\
\beta_c &=& 1.0116 \pm 0.0004
\end{eqnarray}
\end{mathletters}

Monopoles do condense in the confined phase $\beta < \beta_c$. This is
consistent with exact results existing for the Villain action: as explained in
chapter 1 the action in the lattice is determined by requiring that it
coincides with the continuum action in the limit of zero lattice spacing $a\to
0$. This gives a great arbitrariness in terms of higher order in $a$: in theories
like \QCD\   which have a fixed point where $a\to 0$, those terms are expected to
be unimportant. In the language of statistical mechanics they are indeed called
 ``irrelevant'', and models which differ by them are said to belong to the
same class of universality. The phase transition in $U(1)$ is known to be first
order, and therefore strictly speaking these arguments do not apply. Anyhow our
procedure coincides with that of ref.\cite{31} and our result is consistent with
them.

The virtue of the Villain action is that it allows to perform the
transformation to dual variables and to get a lower bound for $\langle\mu\rangle$
for $\beta < \beta_c$. With the generic action such a procedure is not
known: our numerical method supplies to this incovenience.

A duality transformation can also be performed in supersymmetric \QCD\  with
$N=2$, allowing to demonstrate explicitely the condensation of monopoles\cite{34}.

In what follows we shall look for condensation of $U(1)$ monopoles defined by
some abelian projection: of course there we do not know the form of the
effective lagrangean and we only can rely on the numerical methods.

\subsection{Monopole condensation in $SU(2)$, $SU(3)$ gauge theories}
We shall apply the procedure developed in sect.\ref{secu1} for $U(1)$ gauge theory
to $SU(2)$ and $SU(3)$: the extension is kind of trivial, since it consists in
repeating the construction of the $U(1)$ theory to the $U(1)$'s resulting fron the
abelian projection.

We shall restrict our discussion to the abelian projection of the Polyakov loop: in
that case the abelian electric field strength $F_{0i}$ defined by eq.\Ref{eq:v2.6}
only consists of the first term. If $\varphi$ is the Polyakov line, its parallel
transport in the time direction $D_0\varphi = 0$ and
\begin{equation}
F_{0i} = \hat\varphi\cdot\vec G_{0i}\label{eq:6.1b}\end{equation}
The commutation relation between the field strength operators
\begin{equation}
\left[ F_{0i}^a(\vec x,t), F_{0k}^b(\vec y,t)\right] = -\ii
\delta^{ab}\left(\delta_{ij}\partial_k -
\delta_{ik}\partial_j\right)\delta^{(3)}(\vec x - \vec y) + 
\ii 
f^{abc} \left(A^c_k(\vec x)\delta_{ij} - A^c_j(\vec x)\delta_{ik}\right)
\delta^{(3)}(\vec x - \vec y)
\end{equation}
are gauge covariant, and in particular the term proportional to $\delta^{ab}$, i.e.
the commutator for $a=b$, is gauge invariant, and comes from the abelian part of the
field strength.
This implies that the operator constructed in analogy with the $U(1)$ operator
using \Ref{eq:6.1} has magnetic charge $m$. For $SU(2)$ gauge theory in the abelian
projection which diagonalizes the Polyakov line numerical simulation around the
deconfinement phase transition have been performed\cite{35}. As explained in
sect.II this is done by putting the theory on a asymmetric lattice with $N_t \ll
N_s$.

Fig.8 shows the simulation on a $12^3\times 4$ lattices: as
in the $U(1)$ case $\rho$ shows a very sharp negative peak at the deconfining phase
transition. 

The position of the peak agrees with the known value of the deconfining
transition\cite{10} The conclusion is that the $U(1)$ symmetry related to monopole
charge conservation is spontaneously broken, and hence \QCD\ vacuum is indeed a dual
superconductor.

Similar results for the two kinds of monopoles $U(1)\times U(1)$ in the case of
$SU(3)$ are shown in fig.9\cite{36}.

\section{Outlook and future perspectives.}
We have established that \QCD vacuum is a dual superconductor. More studies are
needed to understand what abelian projections define monopole condensing in the
vacuum, and to test 't~Hooft guess that all abelian projections are 
equivalent\cite{28}.

Our method of analysis can also be used to study questions like the order of the
deconfining transition and the possible critical exponents.

In many other numerical investigations\cite{37,38} the number density of monopoles has
been counted across the deconfining phase transition, as well as their contribution to
the observed quantities: the indication is that monopoles do indeed determine the
dynamics (monopole dominance). This information is complementary to our result that
\QCD\ vacuum is a superconductor: it shows that the degrees of freedom involved in
this phenomenon play a dominant role in dynamics.

Discussions with my collaborators, especially with G. Paffuti, and with collegues
T. Suzuki, M. Polikarpov, D. Haymaker, P. Cea, L. Cosmai are aknowledged.

\eject
{\centerline{
{\bf \Large Figure Captions}}}
\begin{itemize}
\item[Fig.1]
 Wilson Loop
\item[Fig.2]
 Spontaneous breaking of $U(1)$ in the Higgs Model.
\item[Fig.3]
 Dirac Monopole.
\item[Fig.4]
 Expected behaviour of $\langle\mu\rangle$ 
in compact $U(1)$ in 4-d.
\item[Fig.5]
 Expected behaviour of $\rho$ in compact $U(1)$ in 4-d.
\item[Fig.6]
$\rho$ for compact $U(1)$ on a $8^3\times16$ lattice.
\item[Fig.7]
Test of the scaling eq.(90). Data come from lattices of different
sizes.
\item[Fig.8]
 $SU(2)$: $\rho$ vs $\beta$ on
a lattice $12^3\times4$
\item[Fig.9]
 $SU(3)$: $\rho$ vs $\beta$ on a lattice $12^3\times 4$.
\end{itemize}
\vfill\eject
{\centerline{\epsfbox{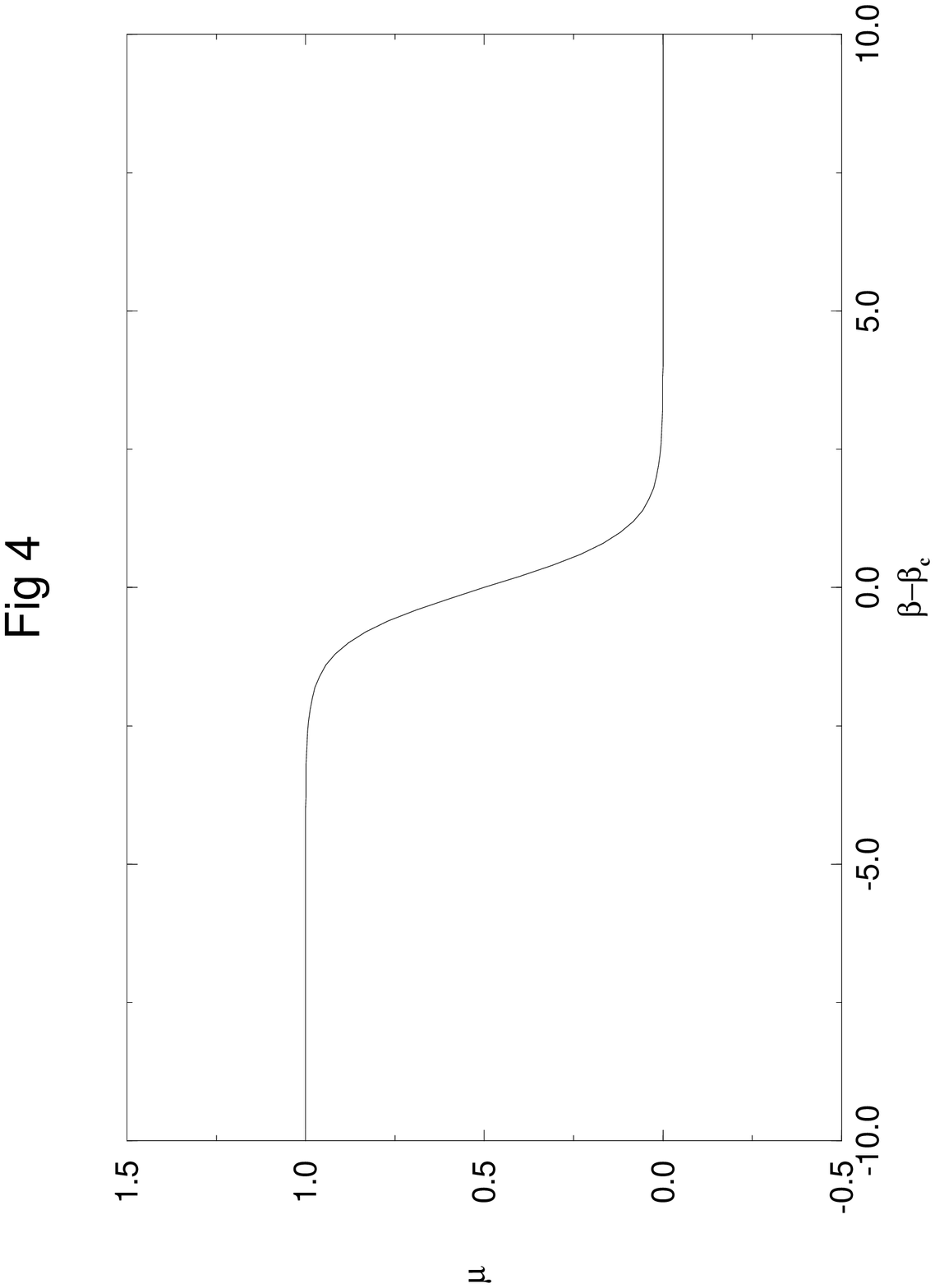}}}
\vfill\eject
{\centerline{\epsfbox{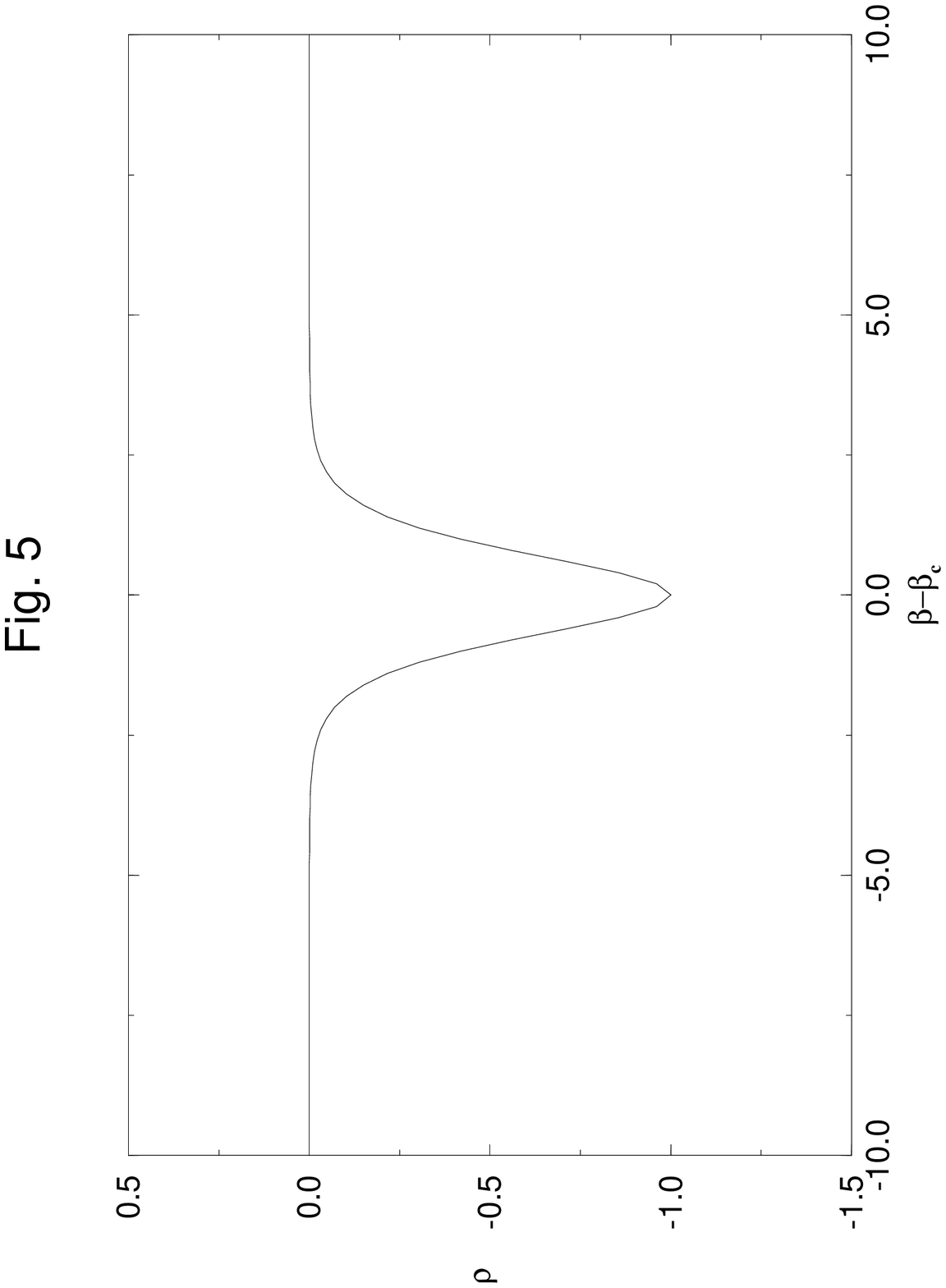}}}
\vfill\eject
{\centerline{\epsfbox{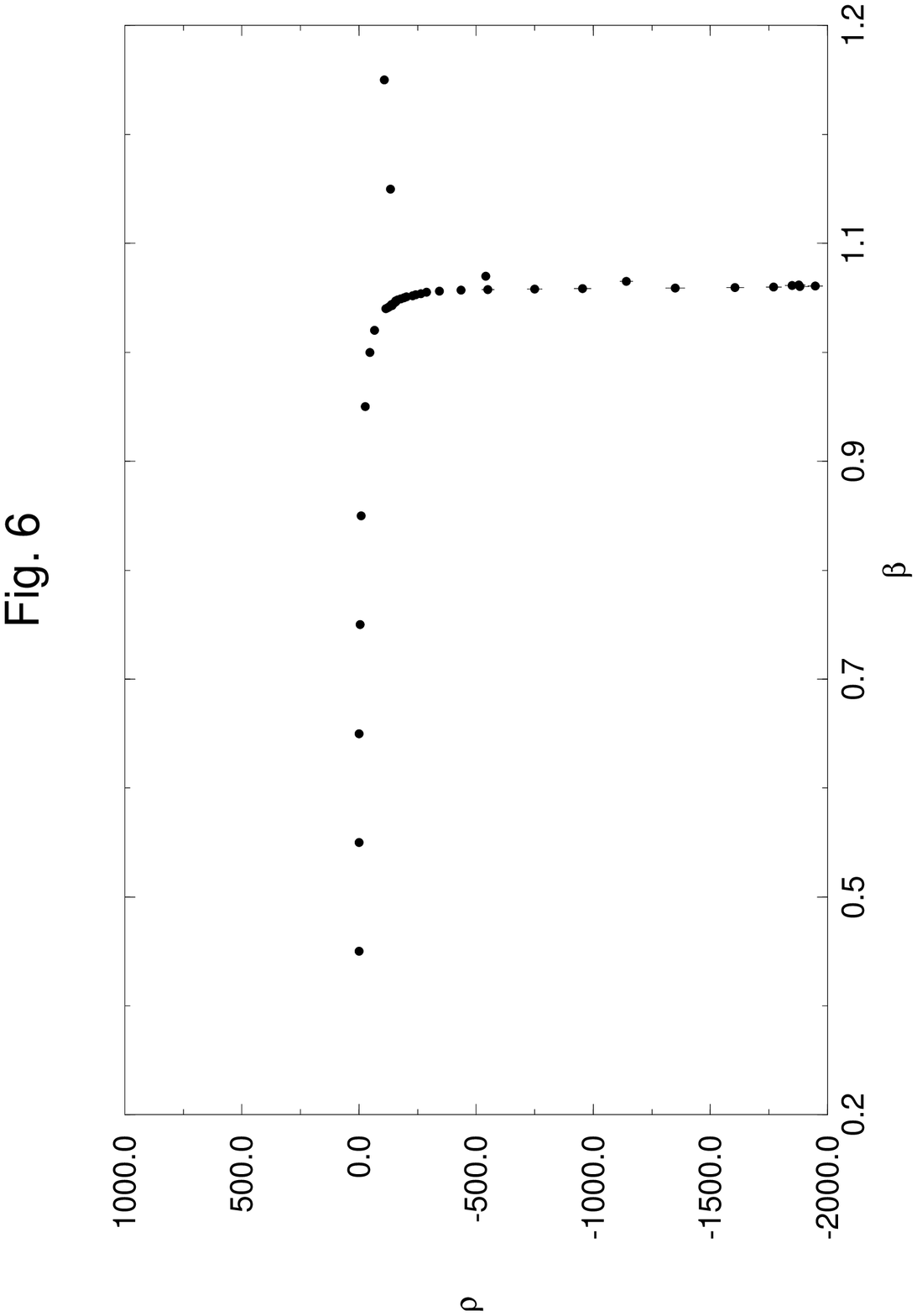}}}
\vfill\eject
{\centerline{\epsfbox{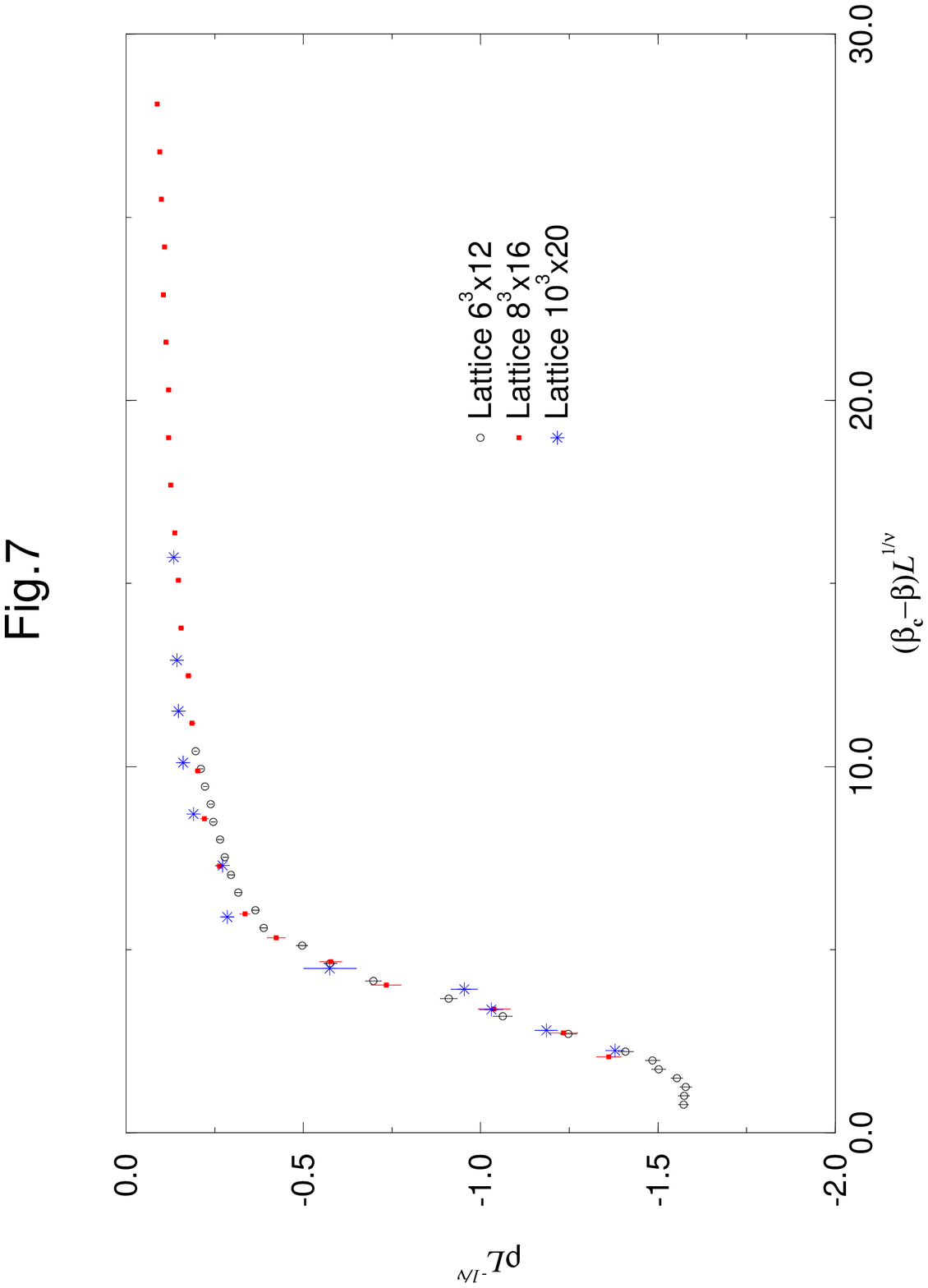}}}
\vfill\eject
{\centerline{\epsfbox{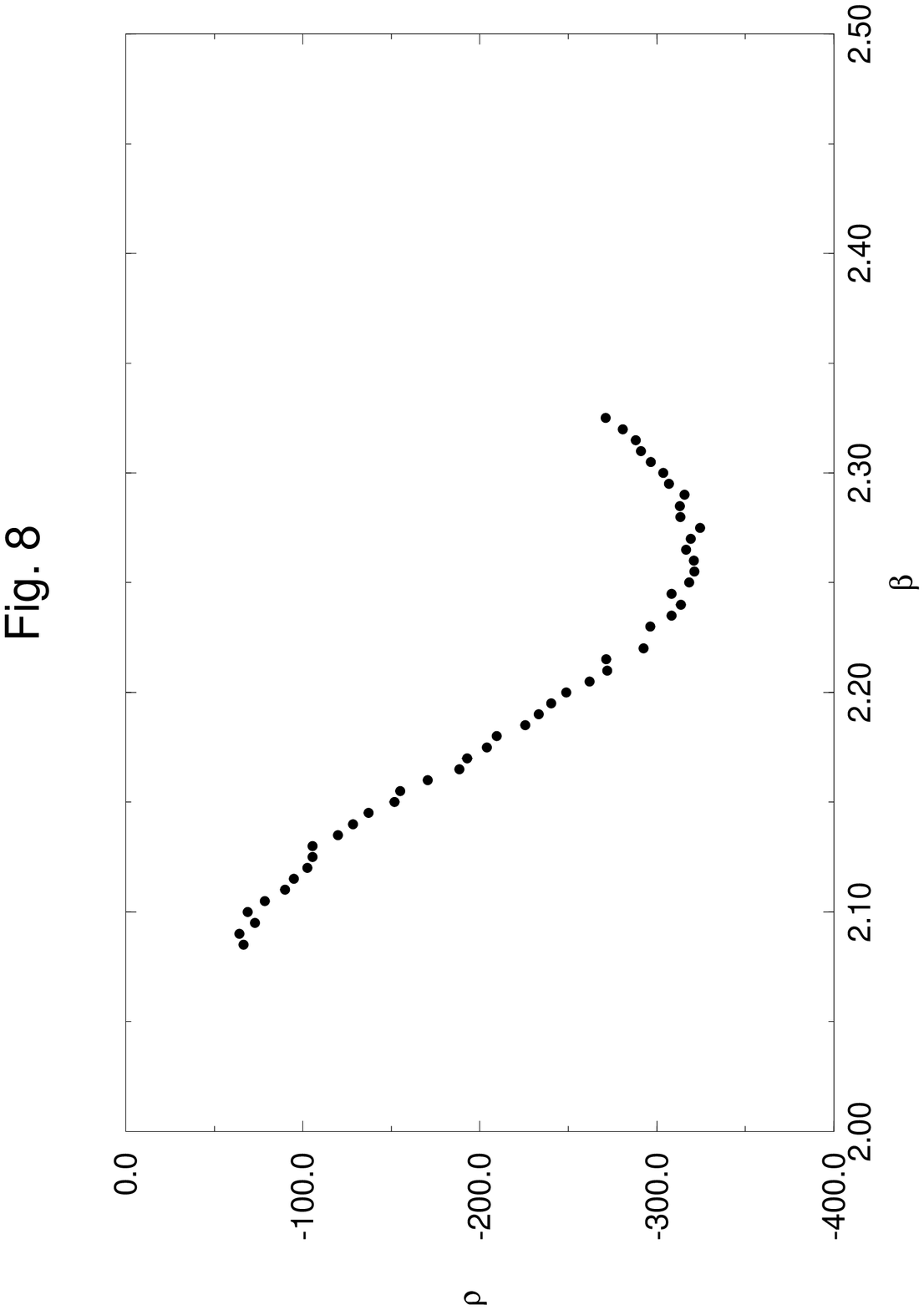}}}
\vfill\eject
{\centerline{\epsfbox{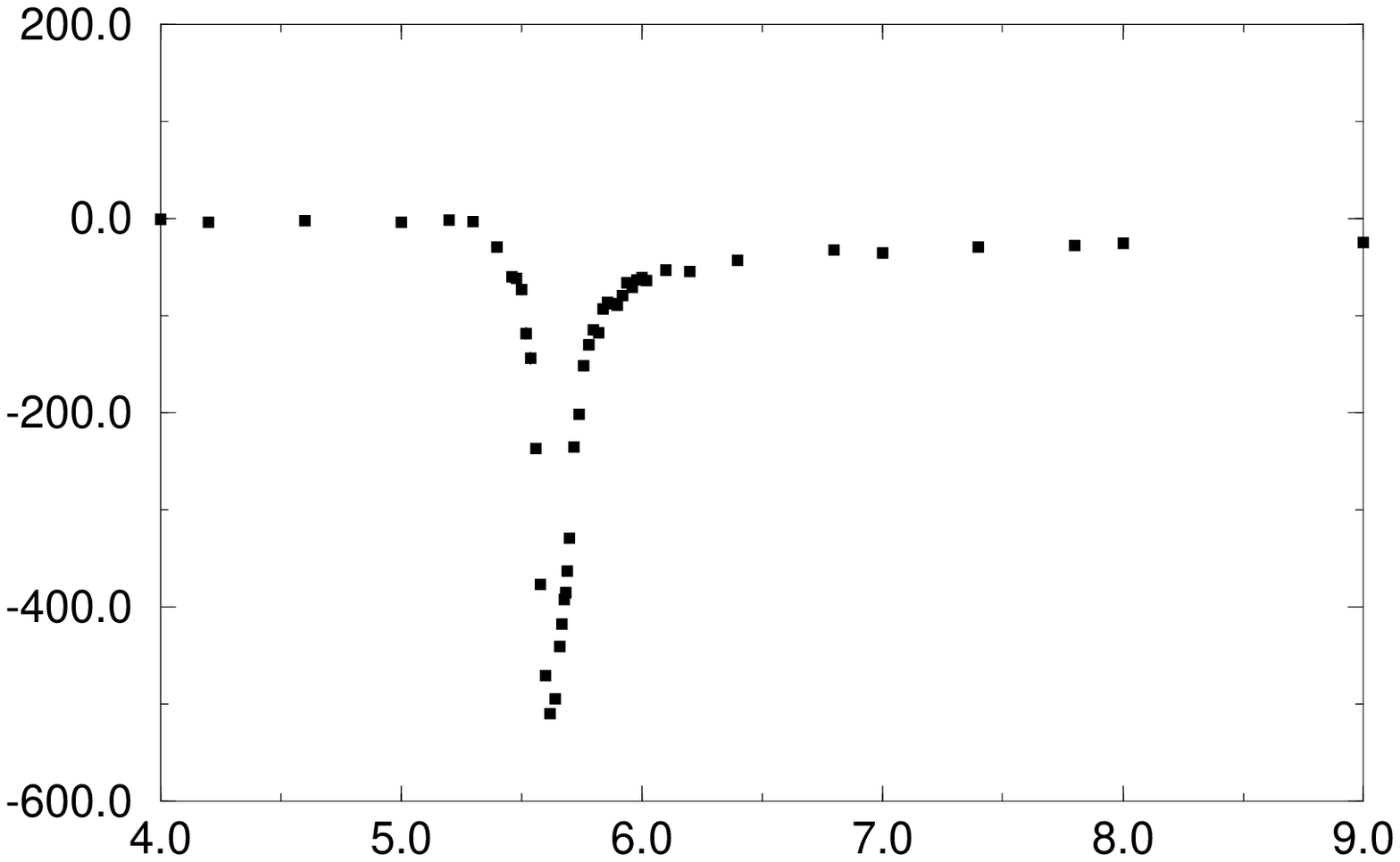}}}
\vfill\eject
\end{document}